\input harvmac
\input epsf
\def\p{\partial}
\def\ap{\alpha'}
\def\half{{1\over 2}}



\Title{}{\vbox{\centerline{A Class of Cosmological Matrix Models }
}}

\centerline{Miao Li}

\centerline{\it Institute of Theoretical Physics}
\centerline{\it
Academia Sinica, P.O. Box 2735}
\centerline{\it Beijing 100080, P.
R. China }
\medskip
\centerline{\it and}
 \centerline{\it Interdisciplinary Center of Theoretical
Studies} \centerline{\it Academia Sinica}

\centerline{\tt mli@itp.ac.cn}

\bigskip

We discuss a class of matrix models describing cosmology with a
light-like singularity, generalizing the model proposed by Craps
et al. in hep-th/0506180.


\Date{June 2005}

\nref\csv{B. Craps, S. Sethi and E. Verlinde,``A Matrix Big Bang,"
hep-th/0506180.}
\nref\hs{G. T. Horowitz and A. Steif, ``Singular string solutions with non-singular
initial data," Phys. Lett. B258 (1991) 91.}
\nref\lms{H. Liu, G. Moore and N. Seiberg, ``Strings in a Time-Dependent Orbifold,"
hep-th/0204168, JHEP 0206 (2002) 045; hep-th/0206182, JHEP 0210 (2002) 031.}
\nref\alor{ A. Lawrence, ``On the instability of 3d null singularities,"
hep-th/0205288, JHEP 0211 (2002) 019.}
\nref\hp{ G. T. Horowitz and J. Polchinski, ``Instability of Spacelike and Null Orbifold
Singularities," hep-th/0206228, Phys.Rev. D66 (2002) 103512.}
\nref\bckr{ M. Berkooz, B. Craps, D. Kutasov and G. Rajesh, ``Comments on Cosmological
Singularities in String Theory," hep-th/0212215, JHEP 0303 (2003) 031.}
\nref\bfss{T. Banks, W. Fischler, S. Shenker and L. Susskind, ``M theory as A Matrix
Model: A Conjecture," hep-th/9610043, Phys. Rev. D55 (1997) 5112.}
\nref\ls{L. Susskind, ``Another Conjecture about Matrix Theory," hep-th/9704080.}
\nref\ns{N. Seiberg, ``Why is the Matrix Model Correct?," hep-th/9710009,
Phys. Rev. Lett.79 (1997) 3577.}
\nref\as{A. Sen, ``D0 Branes on T(n) and Matrix Theory," hep-th/9709220,
Adv. Theor.M ath. Phys. 2 =(1998) 51.}
\nref\wt{W. Taylor IV, ``D-brane Field Theory on Compact Spaces," hep-th/9611042,
Phys. Lett. B394 (1997)283.}
\nref\mr{M. Rozali, ``Matrix Theory and U Duality in Seven Dimensions," hep-th/9702136,
Phys. Lett. B400 (1997) 260.}
\nref\nns{ M. Berkooz, M. Rozali and N. Seiberg, ``Matrix Description of on $T^4$
and $T^5$," hep-th/9704089, Phys. Lett. B408 (1997) 105.}

Craps et al. recently proposed to study a simple cosmological background with a
null singularity \csv, this cosmology admits a matrix model description, thus lends
itself to a rigorous study. An earlier example of cosmology with a null singularity
is proposed in \refs{\hs,\lms} and is later studied by many authors \refs{\alor-\bckr},
they find that this singularity is highly unstable due to gravitational back-reaction.
However, the model of Craps et al. seems to avoid this problem.

The construction of the model in \csv\ is simple. One starts in type IIA string
theory with a flat string metric and a linear dilaton background $\phi=-Qx^+$, where
$x^+$ is a light-like coordinate. The linear dilaton does not require modifying the
critical dimension, since its linearity is along a null direction. Although the string
metric is flat, the Einstein metric is nontrivial:
\eqn\einsm{ds^2=e^{Qx^+/2}(-2dx^+dx^-+(dx^i)^2).}
For a positive $Q$, the metric contracts to a singularity at $x^+=-\infty$, this is
actually a curvature singularity. The corresponding 11 dimensional M theory metric
also has a singularity at $x^+=\infty$. The authors of \csv\ show that the singularity
at $x^+=-\infty$ lies in a finite geodesic distance, while the other singularity
is at infinite distance. This background preserves half of total 32 supersymmetries,
one expects that there is a control over the null singularity.

Since the string metric is flat, this time-dependent background admits a matrix string
description, in which the Yang-Mills coupling constant is time-dependent. In fact,
the authors of \csv\ show that this Yang-Mills theory can be regarded as one with a
constant coupling on a world sheet with a time-dependent metric.

It goes without saying that this is an important observation, and it may provide the
first example in which we can study a time-dependent background in a controlled fashion.
Thus, it is interesting to ask whether this model is unique, or it has many cousins.
In this note we will see that indeed there is a large class of such models.

We shall limit ourselves in M theory in this note. Consider metric
\eqn\mmetric{ds^2=e^{2\alpha x^+}(-2dx^+dx^-+(dx^i)^2)+e^{2\beta x^+}(dx^a)^2,}
where $i=2,\dots 10-d$, $a=11-d,\dots 10$, namely, there are $d$ coordinates
$x^a$, the total dimensions of spacetime is 11. The metric of \csv\ is given by
taking $d=1$, $\alpha=Q/3$ and $\beta=-2Q/3$. $X^{10}$ is taken as the M theory
circle, thus the metric reduces to the IIA flat string metric with a linear
dilaton background.

To find a solution, we use an orthonormal basis
\eqn\ortn{e^\pm=e^{\alpha x^+}dx^\pm, \quad e^i=e^{\alpha x^+}dx^1,
\quad e^a=e^{\beta x^+}dx^a.}
The non-vanishing components of spin connection are
\eqn\spinc{\omega_{+-}=\alpha e^{-\alpha x^+}e^+, \quad \omega_{i+}=\alpha
e^{-\alpha x^+}e^i, \quad \omega_{a+}=\beta e^{-\alpha x^+}e^a.}
The non-vanishing curvature 2-forms are
\eqn\ctwof{\eqalign{R_{i+}&=\alpha^2e^{-2\alpha x^+}e^i\wedge e^+,\cr
R_{a+}&=\beta (2\alpha -\beta) e^{-2\alpha x^+}e^a\wedge e^+.}}
The only non-vanishing component of the Ricci tensor is $R_{++}$ given by
\eqn\riten{R_{++}=[(9-d)\alpha^2+d\beta(2\alpha -\beta)]e^{-2\alpha x^+},}
and the Einstein equation $R_{++}=0$ has two solutions
\eqn\esol{\beta=(1\pm {3\over \sqrt{d}})\alpha.}
For $d=1$, choosing the minus sign in \esol\ reproduces the background considered
in \csv.

We now show that the background given by eqs.\mmetric\ and \esol\ not only is a
solution, but also preserves half of supersymmetries. The only interesting
SUSY transformation is that for the gravitino $\delta \Psi_\mu=D_\mu \epsilon$.
The components of spin connection of interest are $\omega_+$, $\omega_i$ and
$\omega_a$:
\eqn\spinc{\eqalign{\omega_+&=-2\alpha -2\alpha\gamma^-\gamma^+,\quad
\omega_i=2\alpha\gamma^i\gamma^+,\cr
\omega_a&=2\beta e^{(\beta-\alpha) x^+}\gamma^a\gamma^+.}}
Since
\eqn\commc{[D_+,D_a]=\half \beta (\beta-2\alpha)e^{(\beta-\alpha) x^+}\gamma^a
\gamma^+,}
The compatibility condition for the constraints $D_\mu\epsilon =0$ is
$\gamma^+\epsilon=0$. This condition eliminates half of the components in
$\epsilon$. The constraints $D_i\epsilon=D_a\epsilon=0$ are solved if $\epsilon$
is independent of $x^i$ and $x^a$. Finally, $D_+\epsilon=0$ is solved if
$\epsilon=\exp(\half\alpha x^+)\eta$ for a constant $\eta$. We conclude that
the unbroken supersymmetry is parametrized by $\eta$ with the constraint
$\gamma^+\eta=0$. Thus, just like the original background of \csv, our more
general metric also preserves 16 supersymmtries, this should be enough to
guarantee that our proposed matrix model to be described shortly is a valid
description of dynamics over this background.

In the case $d=9$, $\beta=0, 2\alpha$, although $[D_+,D_a]=0$, but
$[D_+,D_i]=-\half\alpha^2\gamma^i\gamma^+$, again we obtain the same unbroken
supersymmetries.

In fact, \mmetric\ is only a special case of a more general class of solutions
preserving half of supersymmetries (to be discussed in the end of this note).
Before we study the matrix model for \mmetric, let us discuss the geometric
properties of \mmetric. Apparently, the sign of $\alpha$ and its absolute
value can be changed by changing $x^+$ and $x^-$, so we always assume $\alpha
>0$. As $x^+\rightarrow -\infty$, the factor $e^{2\alpha x^+}$ approaches zero,
the transverse dimensions $x^i$ shrink to zero size, this is the big bang point with
regard to these coordinates. This singularity locates a finite distance away
in view of the affine parameter, indeed, let $dX^+=e^{2\alpha x^+}dx^+$, the first
term in the metric \mmetric\ becomes
\eqn\lightm{-2dX^+dx^-,}
thus $X^+$ is the affine parameter for the null geodesic $x^-=$const. The big bang
singularity occurs at $X^+=0$.

The geometry of other transverse dimensions $x^a$ depends on the choice $\beta$
in \esol. We name the choice $\beta=(1+3/\sqrt{d})\alpha$ case 1, and the
choice $\beta=(1-3/\sqrt{d})\alpha$ case 2. For case 1, $\beta>0$, so all $x^a$
shrink at $x^+=-\infty$. For case 2, $\beta <0$ except for the $d=9$ case, thus
all $x^a$ get to infinity at $x^+=-\infty$ and shrink to zero size at $x^+=\infty$.
$d=9$ is special, in this case, all 9 dimensions $x^a$ do not evolve in time, and one
can redefine $x^+$ such that the metric is Minkowski. Supersymmetry is also enhanced,
since there is no longer constraint $\gamma^+\epsilon=0$ following from \commc.

The model studied in \csv\ is a special example of case 2 when $d=1$. As in \csv,
we can compactify $x^{10}$ on a circle to obtain type IIA string theory. The string
metric and the dilaton $\phi$ are related to the M theory metric through
\eqn\stmr{ds^2=e^{-2\phi/3}ds_{st}^2+e^{4\phi/3}(dx^{10})^2,}
we obtain
\eqn\dstm{\phi={3\over 2}\beta x^+,\quad ds_{st}^2=e^{(2\alpha+\beta )x^+}
[-2dx^+dx^-+(dx^i)^2]+e^{3\beta x^+}(dx^a)^2.}
The 10D Einstein metric reads
\eqn\tend{ds_E^2=e^{(2\alpha +\beta/4)x^+}[-2dx^+dx^-+(dx^i)^2]+e^{(9\beta/4)x^+}
(dx^a)^2.}
Since the string coupling ``constant" $g_s=\exp(\phi)=\exp({3\over 2}\beta x^+)$,
for case 1, strings are weakly coupled at big bang, while all transverse dimensions
are zero-sized, so it is not clear whether we can trust the perturbative string theory.
As we shall see shortly in the matrix model, spacetime breaks down and we need to employ
a full non-abelian description. At later times, strings become strongly coupled,
while all dimensions expand, we shall see that a rather simple theory emerges, that is,
only abelian degrees of freedom survive.

For case 2, strings are strongly coupled at big bang. transverse dimensions $x^i$
start with a zero size, while transverse dimensions $x^a$ have infinite size and contract
as time evolves. This picture in valid both in terms of the string metric as well as
the 10D Einstein metric. We will study in more detail the spacetime properties in the matrix
model later.

The string frame metric \dstm\ is in general non-flat. $\beta$ never vanishes, so to get
a flat metric, $d=1$ is necessary, and in this case one chooses $\beta=-2\alpha$, this is
the special case considered in \csv. More generally, one obtains a world-sheet theory
with an action explicitly depending on time. One can attempt to quantize the string in
the light-cone gauge, to do so, introduce the new light-cone coordinate $dy^+=\exp((2\alpha
+\beta)x^+)dx^+$, the first term in the string frame metric \dstm\ becomes
$-2dy^+dx^-$. The lignt-cone momentum $p_-=-p^+$ conjugate to $x^-$ is conserved.
In the light-cone quantization, we can set the light-cone gauge $y^+=\tau$, the bosonic part
of the world-sheet action reads
\eqn\wsa{S={1\over 4\pi\ap}\int d\tau d\sigma [e^{(2\alpha+\beta)x^+}\p^\alpha X^i\p_\alpha
X^i+e^{3\beta x^+}\p^\alpha X^a\p_\alpha X^a],}
where the period of $\sigma$ is $2\pi\ap p^+$, and $\exp((2\alpha+\beta)x^+)=(2\alpha+\beta)
\tau$. As $x^+\rightarrow -\infty$, $\tau\rightarrow 0$. Since $\tau$ starts at a finite
point, it is more useful to use the old light-cone coordinate $x^+=t$, the action becomes
\eqn\wsar{S={1\over 4\pi\ap}\int dt d\sigma [e^{(4\alpha+2\beta)t}\p^\alpha X^i\p_\alpha
X^i+e^{(2\alpha+4\beta)t}\p^\alpha X^a\p_\alpha X^a].}
The time-dependent coefficients may be interpreted as the effective tensions. For transverse
coordinates $X^i$, the effective tension is $T_i={1\over 2\pi\ap}e^{(4\alpha+2\beta)t}$, for
coordinates $X^a$, the effective tension is $T_a={1\over 2\pi\ap}e^{(2\alpha+4\beta)t}$.
In case 1, both effective tensions get to zero at big bang, while the string coupling
also gets to zero. Since the string spectrum becomes very dense, the usual free string picture
does not apply. In later times, effective string tensions get large, and the string coupling
constant also becomes strong, we will have to deal with a strongly coupled massless sector.
In case 2, $4\alpha +2\beta>0$ except for $d=1$, our above analysis still applies to $T_i$.
$2\alpha +4\beta=6\alpha (1-2/\sqrt{d})$, for $d>4$, $T_a$ becomes small in earlier time,
the same as case 1. For $d<4$, $T_a$ becomes large in earlier time, thus these transeverse
dimensions are effectively frozen. $d=4$ is a special case, $T_a$ is independent of time.

The vertex operator of a string state in the background \dstm\ in general is quite complicated.
For instance, consider a massless scalar satisfying equation of motion
\eqn\malsca{\p_\mu \left(e^{-2\phi}\sqrt{-g}g^{\mu\nu}\p_\nu\Phi\right)=0.}
As function of $x^+$, the scaling of components $g^{ab}$ is different from the scaling of
components $g^{+-}$ and $g^{ij}$, so there is no simple plane wave solution in general, which
implies that the vertex operator of this massless scalar field is not simple. However, we can
consider special cases when $\Phi$ is independent of $x^a$, in this case, a vertex operator
$V=\exp(ik_+x^++ik_-x^-+ik_ix^i)$ must satisfy the on-shell condition:
\eqn\onsh{k_-(2k_+-i\gamma)-k_i^2=0,\qquad \gamma=(9-d)\alpha+d\beta,}
which require a imaginary part of $k_+$: $\Im k_- =\gamma/2$, thus in the vertex operator, there
is an exponential factor
\eqn\exppp{e^{-\half \gamma x^+}.}
Since $\gamma=(9\pm 3\sqrt{d})\alpha\ge 0$, this exponential factor always blows up at $x^+=
-\infty$ for $d<9$. The effective string coupling constant for these states is $g_{eff}=g_s\exp(-\half
\gamma x^+)$. Dimensions $x^a$ for these states are effectively compactified, thus when we
discuss interactions among these states, the space-time dimensionality is $11-d$, and
there is an effective string tension in action
\wsar, or $\ap_{eff}=\ap \exp(-(4\alpha+2\beta)x^+)$. The effective Newton constant is
$G_{eff}=g_{eff}^2\ap^{(9-d)/2}$, as a function of $x^+$, it scales
\eqn\effn{G_{eff}\sim e^{-\gamma x^+-(9-d)(2\alpha +\beta)x^+}.}
Now, $\gamma$ is positive and $(9-d)(2\alpha+\beta)$ is non-negative if $d<9$, the effective Newton
constant blows up at $x^+=-\infty$, the perturbative sstring picture is not valid at least
for these states.

Next, we study the matrix model. It is not necessary to compactify any dimension of \mmetric.
In the usual flat background, the Matrix Theory action reads \bfss
\eqn\mact{S=\int dt\Tr\left({1\over 2R}(D_tX^i)^2+{R\over 4}[X^i,X^j]^2+i\theta^TD_t\theta
-R\theta^T\gamma_i[X^i,\theta]\right),}
where $R$ is the longitudinal cut-off, or it may be viewed as the radius of $x^-$ in the
DLCQ M theory \ls. For simplicity, we set the M theory Planck length $l_p=1$. A derivation
of this matrix action is given in \ns\ (see also \as\ ).
With our metric \mmetric, it is straightforward to write the corresponding matrix action.
To do this, we need to use the light-cone coordinate in \lightm, and identify $X^+$ with
$\tau$, the world-line time of D0-branes.  In the matrix model, we apply the M theory metric
\mmetric\ directly. Since the action is rather lengthy, we separate the action into the
bosonic part and the fermionic part. The bosonic part reads
\eqn\bma{\eqalign{S_B&=\int d\tau\Tr\{{1\over 2R}e^{2\alpha x^+}(D_\tau X^i)^2+{1\over 2R}
e^{2\beta x^+}(D_\tau X^a)^2+{R\over 4}e^{4\alpha x^+}[X^i,X^j]^4\cr
&+{R\over 4}e^{4\beta x^+}[X^a,X^b]^4+{R\over 2}e^{(2\alpha+2\beta) x^+}[X^i,X^a]^4\}.}}
Note that $x^+$ appearing in the above action is the old light-cone coordinate. The fermionic
part reads
\eqn\fma{S_F=\int d\tau\{i\theta^TD_\tau\theta -Re^{\alpha x^+}\theta^T\gamma_i
[X^i, \theta ]-Re^{\beta x^+}\theta^T\gamma_a [X^a, \theta ]\}.}

It is rather awkward to use $\tau$ as time, since it has a finite beginning. Let us switch
back to the coordinate $x^+$ and on the world-line identify $t=x^+$, thus $d\tau
=\exp(2\alpha t)dt$, we have
\eqn\rbma{\eqalign{S_B&=\int dt\Tr\{{1\over 2R}(D_t X^i)^2+{1\over 2R}
e^{2(\beta-\alpha) t}(D_t X^a)^2+{R\over 4}e^{6\alpha t}[X^i,X^j]^4\cr
&+{R\over 4}e^{(2\alpha+4\beta)t}[X^a,X^b]^4+{R\over 2}e^{(4\alpha+2\beta) t}[X^i,X^a]^4\}
,}}
and
\eqn\rfma{S_F=\int dt\{i\theta^TD_t\theta -Re^{3\alpha t}\theta^T\gamma_i
[X^i, \theta ]-Re^{(2\alpha+\beta)t}\theta^T\gamma_a [X^a, \theta ]\}.}

Before study the simplest properties of this matrix model, let us show that we
can recover the matrix string action of \csv. In this case, there is only one
$X^a$, call it $X^{10}$. The compactification scheme is given in \wt.
Up a dimensionful parameter, we replace the trace
in the matrix action by $\int d\sigma \Tr$, the commutator $R[X^{10},X^i]$
is replaced by the covariant derivative $iD_\sigma X^i$. Finally, use $\beta
=-2\alpha$, we find
\eqn\bmsa{S_B=\int dtd\sigma\Tr\{{1\over 2R}(D_\alpha X^i)^2+{1\over 2R^3}g_s^2
F_{t\sigma}^2+{R\over 4}g_s^{-2}[X^i,X^j]^2\},}
and
\eqn\fmsa{S_F=\int dt d\sigma\Tr [\theta^T\sigma^\alpha D_\alpha\theta -
Rg_s^{-1}\theta^T\gamma_i[X^i,\theta ],}
where $g_s$ is the time-dependent string coupling $g_s=\exp(-3\alpha t)$. Our action
is identical to that in \csv\ up to an identification of a dimensionful parameter
$l_s$. This matrix string theory can be regarded as a 2D Yang-Mills theory with a
time-dependent coupling or a 2D Yang-Mills theory with a constant coupling in the
world-sheet metric
\eqn\wsm{ds^2=g_s^{-2}(-dt^2+d\sigma^2).}

To study the properties of matrix model defined by \rbma\ and \rfma, we consider
cases 1 and 2 separately.

$\bullet$ Case 1, $\beta=(1+{3\over \sqrt{d}})\alpha$.

The kinetic term of $X^i$ is always simple, as is the kinetic term of $\theta$.
The coefficients of the remaining terms all vanish in the limit $t\rightarrow
-\infty$, so there is no constraint arising from these terms, this implies
that all matrices are fully non-abelian. On the other hand, as $t\rightarrow \infty$,
these coefficients blow up, thus, all matrices must commute with one another,
the only surviving degrees of freedom are diagonal elements. Moreover, $X^a$ must
be independent of time in this limit, so $X^a$ become frozen abelian moduli. Recall
that in the string picture, we found previously that the effective tension $T_a$
becomes infinitely heavy, even heavier than $T_i$ for large $t$, this is related to
the fact that $X^a$ become moduli in the matrix model, while $X^i$ still have dynamics.
If we compactify some of the transverse
dimensions, the story becomes slightly more involved. For instance, compactifying
$X^{10}$ on a circle, all other matrices become function on a circle $\sigma$.
In the limit $t\rightarrow \infty$, although they have to be diagonal, they are
not always periodic functions of $\sigma$, the eigen-values can get permuted
after circling along $\sigma$, these twisted sectors describe strings of various
lengths.

$\bullet$ Case 2, $\beta=(1-{3\over \sqrt{d}})\alpha$.

$6\alpha$ and $4\alpha+2\beta$ are always positive, at big bang, there is
no constraint on commutators $[X^i,X^j]$, $[X^i,X^a]$, $[X^i,\theta ]$
and $[X^a,\theta]$. On the other hand, as $t\rightarrow \infty$, these
commutators are forced to vanish. $2(\beta-\alpha)$ is always negative,
so at bing bang, $X^a$ must be independent of time, they become non-abelian
moduli in the model. $2\alpha -4\beta=6\alpha (1-2/\sqrt{d})$, $d=4$ becomes
the critical dimension. For a larger $d$, there is no constraint on the
commutators $[X^a,X^b]$ at big bang, and they have to vanish as $t\rightarrow
\infty$, thus, for $d>4$, all matrices commute in this limit and only
abelian degrees of freedom survive. For $d<4$, $[X^a,X^b]$ have to vanish
at big bang, together with the fact that $X^a$ are independent of time,
these matrices become abelian moduli at big bang, this fact is also
reflected in our previous analysis in the string picture, where we found
$T_a$ become infnitely heavy at big bang. When $d=4$, $X^a$ remain
nonabelian all the time. In the string picture, recall that when $d=4$, the
effective tension $T_a$ is constant.

It is also of interest to study compactification. Just as in the flat background,
we do not know how to write down matrix model action if we compactify more than
5 dimensions. We may choose to compactify all $X^a$, or some of $X^a$, or some
of $X^a$ and some of $X^i$. For simplicity, let us compactify all $X^a$ on a
torus $T^d$. Replacing
$R[X^a,X^i]$ by $iD_a X^i$, $R^2[X^a, X^d]$ by $-F_{ab}$ etc, the bosonic part of
the matrix model action reads
\eqn\cbma{\eqalign{S_B&=\int d^{d+1}\sigma\{{1\over 2R}(D_tX^i)^2-{1\over 2R}
e^{(4\alpha+2\beta)t}(D_aX^i)^2+{1\over 2R^3}e^{2(\beta-2\alpha)t}F_{ta}^2\cr
&+{1\over 4R}e^{(2\alpha+4\beta)t}F_{ab}^2+{R\over 4}e^{6\alpha t}[X^i,X^j]^2\}.}}
The analysis of this action is similar to our previous analysis of the action
without compactification, a statement there can be simply translated to one for
the action \cbma. For instance, demanding $[X^a,X^b]=0$ is translated to $F_{ab}
=0$, that is, the spatial connection must be flat.

It is interesting that the previous noticed ``critical dimension" $d=4$ corresponds
to the situation that the action \cbma\ is no longer complete. Compactification on
a torus $T^4$, the complete matrix model is to be given by the world volume theory of
coincident M5-branes \mr. Of course, that theory will also be time-dependent,
and hopefully one can figure
out more details of the time dependence by studying field theory behavior of \cbma.
Likewise, action \cbma\ is again incomplete for compactification on $T^5$, in this
case the complete theory is given by the little string theory \nns.

Unlike the case when $d=1$, the Yang-Mills theory \cbma\ can not be simply interpreted
as a theory with a constant coupling on a world volume of a nontrivial metric. To
interpret action \cbma\ as Yang-Mills theory, we need to introduce both a nontrivial
metric on the world volume as well as a time-dependent Yang-Mills coupling.
Since a factor $g_{YM}^{-2}$ appears in the coefficient of $F_{\mu\nu}^2$ and a factor
$g_{YM}^2$ appears in the coefficient of $[X^i,X^j]^2$, we find for $d>1$
\eqn\mcou{\eqalign{ds_{WV}^2&=-e^{{d\over d-1}(4\alpha+2\beta)t}dt^2+e^{{1\over d-1}
(4\alpha+2\beta)t}(d\sigma^a)^2,\cr
g_{YM}^2&=e^{-{1\over d-1}(4\alpha+2\beta)t+2(\alpha-\beta)t},}}
where the subscript $WV$ stands for world volume. the world volume metric components $g_{aa}$
must be all same, there are only 3 independent functions in the geometry and the Yang-Mills
coupling, however, we need to match 5 different kinds of coefficients in \cbma, so
it is nontrivial that there is a solution as presented in \mcou. This is quite important especially
for $d>3$, since we need to complete the action \cbma\ by introducing more degrees
of freedom, this is possible only when \cbma\ admits Yang-Mills theory interpretation.
One can check that the identification \mcou\ also works for the fermionic part
of the matrix action.

Finally, we discuss generalizations of background \mmetric. There are two directions
to generalize \mmetric. The first is to consider
\eqn\mmet{ds^2=-2e^{2\alpha x^+}dx^+dx^-+\sum_ie^{2\beta_i x^+}(dx^i)^2.}
The Einstein equations are solved provided
\eqn\aleq{\alpha^2+\sum_i \beta_i(2\alpha -\beta_i)=0.}
This background again preserves 16 supersymmetries of the form $\epsilon=
\exp(\half \alpha x^+)\eta$, with a constant $\eta$ satisfying $\gamma^+\eta=0$.

The second direction to generalize \mmetric\ is to consider metric of the form
\eqn\mmmet{ds^2=e^{2f(x^+)}(-2dx^+dx^-+(dx^i)^2)+e^{2g(x^+)}(dx^a)^2.}
The non-vanishing curvature 2 forms are
\eqn\ctwof{\eqalign{R_{i+}&=(f'^2-f'')e^{-2f}e^i\wedge e^+,\cr
R_{a+}&=(2f'g'-g'^2-g'')e^{-2f}e^a\wedge e^+.}}
The Einstein equations are solved if
\eqn\dfeq{(9-d)(f'^2-f'')+d(2f'g'-g'^2-g'')=0.}
Once again, there are 16 unbroken supersymmetries, parameterized by
$\epsilon =\exp(\half f)\eta$, $\gamma^+\eta=0$.

Needless to say that
equation \dfeq\ admits infinite many solutions. Two special solutions
deserve attention. One solution is when $g=0$, in this case $f'^2-f''=0$.
All the curvature 2 forms vanish thus it may appear that we obtain a flat space
solution. This is almost case except for a simple twist. The solution
of $f'^2-f''=0$ is $\exp(2f)=1/(x^+)^2$, by changing coordinates $dy^+=dx^+/(x^+)^2$,
$y^-=x^-$ we obtain a metric
\eqn\orbm{ds^2=-2dy^+dy^-+(y^+)^2(dx^i)^2+(dx^a)^2.}
If one of $x^i$ is periodic, we find a nontrivial background similar to the orbifold
discussed in \refs{\hs, \lms}. Of course if all $x^i$ are noncompact, \orbm\
is equivalent to the Minkowski space. Similarly, one may set $f=0$ and find
$\exp(2g)=(x^+)^2$, a nontrivial orbifold results provided one of $x^a$
is compactified. One may attempt to construct a matrix model for the metric
\orbm\ too. However, time $\tau=y^+$ seems to have to terminate at $\tau=0$.
One may try to avoid this problem by going back to the original light-cone
coordinate $x^+$ and take $t=x^+$ in the matrix model, although this
helps to eliminate the coefficient $\tau^2$ in the kinetic term of $X^i$ but
it also introduces a coefficient $t^2$ in the kinetic term of $X^a$, again
$t$ must be terminated at $t=0$. Also the supersymmetry parameter $\epsilon
=\sqrt{x^+}\eta$ does not exist beyond $x^+=0$ since $\epsilon$ has to be
real. This problem may cause disease in the definition of the matrix model, and
may be a reflection of the gravitational instability caused by a test particle.

Acknowledgments.

I am grateful to B. Chen, Q. G. Huang and W. Song for discussions.
This work was supported by a grant of CNSF.

\listrefs

\end